# Surface termination effect of SrTiO$_3$ substrate on ultrathin SrRuO$_3$

Huiyu Wang[1], Zhen Wang[1,2,4]*, Zeeshan Ali[2], Enling Wang[3], Mohammad Saghayezhian[3], Jiandong Guo[3], Yimei Zhu[4]*, Jing Tao[1] and Jiandi Zhang[3,2]*

[1]*Department of Physics, University of Science and Technology of China, Hefei, Anhui 230026, People's Republic of China*

[2]*Department of Physics & Astronomy, Louisiana State University, Baton Rouge, LA 70803, USA.*

[3]*Beijing National Laboratory for Condensed Matter Physics, Institute of Physics, Chinese Academy of Sciences, Beijing 100190, People's Republic of China*

[4]*Condensed Matter Physics & Materials Science, Department, Brookhaven National Laboratory, Upton, NY 11973, USA.*

*Corresponding authors: wangzhen03@ustc.edu.cn; zhu@bnl.gov; jiandiz@iphy.ac.cn

**ABSTRACT**

A uniform one-unit-cell-high step on the SrTiO$_3$ substrate is a prerequisite for growing high-quality epitaxial oxide heterostructures. However, it is inevitable that defects induced by mixed substrate surface termination exist at the interface, significantly impacting the properties of ultrathin films. In this study, we microscopically identify the origin for the lateral inhomogeneity in the growth of ultrathin SrRuO$_3$ films due to the step effects of SrTiO$_3$(001). By using atomic-resolved scanning transmission electron microscopy, we observe two distinct types of step propagation along the [011] and [0$\bar{1}$1] crystallographic direction in SrTiO$_3$-SrRuO$_3$ heterostructures, respectively. In particular, the type-II [0$\bar{1}$1] step results in lateral discontinuity of monolayer SrRuO$_3$ and originates from the SrO-terminated regions along the TiO$_2$-terminated step edge. Such an induced lateral discontinuity should be responsible for the distinct electronic and magnetic properties of monolayer SrRuO$_3$. Our findings underscore the critical importance of using single termination STO substrate to achieve high-quality termination selective films and to unveil the intrinsic properties of epitaxial films in the atomic limit.

# I. INTRODUCTION

SrTiO$_3$(STO) is a popular substrate because of its compatible lattice parameters with many perovskite oxides. It is the most commonly used substrate for the growth of functional materials, such as high-temperature superconducting, colossal magneto-resistive, and multiferroic films [1–4]. A distinctive characteristic of the STO substrate is the atomic control of a terrace-like structure on its surface after chemical and thermal treatment [5–7]. The morphology of substrate-surface step is a sensitive factor that affects the film growth, microstructures, and consequently physical properties. First, uniform unit-cell height steps on the substrate can effectively prevent the formation of three-dimensional islands during film growth, which is beneficial and a prerequisite for realizing high-quality epitaxial films [8,9]. Second, vicinal STO substrates can be used to manipulate film properties by introducing defective structures in thick films, such as antiphase domains [10], out-of-phase domains [11], and stacking faults [12], by taking advantage of step-induced out-of-plane lattice mismatch, structural domain variants regulated by terrace facets [13], and dislocations by providing preferential nucleation sites [14]. These defective structures can lead to ferroelectric and ferromagnetic domains, and act as flux and free electron pinning centers [11,15]. Thirdly, steps on the STO substrate have a profound effect on the physical properties of interfaces and ultrathin films. Surface measurements indicate that the presence of steps can lead to an uneven distribution of electrostatic potential [16], which will inevitably lead to inhomogeneity of crystal structure and electronic structure at interfaces and within ultrathin films. For instance, steps induced resistivity anisotropy in LAO/STO [17], which can be attributed to a deteriorative effect on the conductive properties of interfaces [18]. Therefore, revealing the impact of steps and terraces on the atomic and electronic structures of interfaces and ultrathin films is essential for understanding their structure-property relationships.

Current research primarily relies on surface measurements to explore ultrathin film properties and to deduce possible structural and electronic reconstructions that occur at the steps. However, detailed atomic-scale studies of step-induced microstructures using scanning transmission electron microscopy (STEM) are lacking. The main challenge is pinpointing the location of steps at the substrate-film interface and tracking step propagation within the grown film. Two factors make this challenging: firstly, substrate steps do not introduce obvious structural defects in ultrathin films, such as domains and dislocations. Secondly, the STEM sample must be cut perpendicular to the step, as any overlap in the projected direction will blur the precise position of the step. Here in this work, we address this challenge and focus on studying the step-induced microstructures in ultrathin SrRuO$_3$(SRO) superlattices grown on STO substrate.

Bulk 4d transition-metal oxide SRO exhibits a ferromagnetic (FM)-metallic ground state with a Curie temperature $T_C \sim 160$ K [19]. The metal-insulator transition (MIT) in SRO thin films has been reported at a critical thickness ranging from 4-5 unit cells (u.c.) to 2 u. c. [20–24]. The thickness-dependent MIT transition is attributed to factors such as enhanced electronic correlations [19–21], structural transition [28], dynamic spin correlations [29], and extrinsic effects such as surface disorder and nonstoichiometry [22,23]. Moreover, the ground states of monolayer SRO superlattices, which remove the surface disorder effect, are obtained from a non-FM insulator [23,24] to an FM insulator [31] to borderline FM metal [25]. Notably, these experimental results contradict the theoretically proposed FM half-metallic state for monolayer SRO confined within the STO lattice [20]. Excluding extrinsic factors will help us to address the inconsistences between experimental and theoretical results and uncover the intrinsic properties of monolayer SRO. One crucial question is whether the commonly observed substrate-surface steps result in defects in monolayer SRO. In addition, how the step propagates in the SRO film, which would provide direct evidence for the film growth mode, remains unclear thus far.

To address these questions, we studied the step-induced microstructures in STO$^5$-SRO$^2$-STO$^5$-SRO$^1$ heterostructure grown on a vicinal STO(001) substrate using atomic-resolved STEM and electron energy loss spectroscopy(EELS). Unlike atomic force microscopy (AFM) and scanning tunneling microscopy (STM), which are commonly utilized to investigate the SRO growth process through the surface morphology, our STEM observations directly provide atomic structure information in the cross-sectional view of the interface and film. We observed two types of step propagation modes in the film specifically along the [011] and [0$\bar{1}$1] crystallographic directions, which have distinct effects on the lateral homogeneity of the SRO layers. The step propagation mode is found to be related to the SrO-terminated region along the step edge. Our findings provide new insight into the mechanism underlying the physical properties of interfaces and ultrathin films.

## II. METHODS

### A. Film growth and surface characterization

Heterostructures of $STO^5$-$SRO^2$-$STO^5$-$SRO^1$ and $STO^5$-$SRO^3$-$STO^5$-$SRO^1$ were fabricated via pulsed-laser deposition (PLD) on $TiO_2$ terminated STO (001) substrates. The STO substrates were first sonicated in deionized water and then treated for 30 s in buffered hydrogen fluoride (BHF), and finally annealed at 950 °C in an oxygen atmosphere to generate atomically smooth surfaces [26]. The $TiO_2$-terminated STO (100) substrate is atomically smooth with steps of one unit cell in height. The SRO and STO films were grown at 650 °C with oxygen pressures of 100 and 10 mTorr, respectively. A KrF excimer laser ($\lambda$ = 248 nm) laser repetition with a rate of 10 Hz (SRO) and 5 Hz (STO), and energy of 300 mJ (SRO) and 260 mJ (STO) was used. After deposition, the samples were cooled at ∼12°/min to room temperature in 100 mTorr oxygen. The surfaces of the substrates were characterized using an atomic force microscope (AFM). The film thickness was monitored by *in situ* reflection high-energy electron diffraction (RHEED).

### B. Composition and structural characterization

TEM samples were prepared using a focused ion beam with Ga+ ions followed by Ar+ ion milling to a thickness of ∼30 nm. These TEM samples were cut in the direction perpendicular to the steps to ensure that the probing electron beam was parallel to the step edge without atom overlap. STEM and EELS experiments were conducted using a 200-kV JEOL ARM electron microscope equipped with double-aberration correctors, a dual energy-loss spectrometer, and a cold field-emission source. The atomic-resolution high-angle annular dark-field (HAADF) STEM image was collected with a 21-mrad convergent angle and a collection angle of 67–275 mrad. The microscope conditions were optimized for EELS acquisition with a probe size of 0.8 Å, a convergence semi-angle of 20 mrad, and a collection semi-angle of 88 mrad. EELS mapping was obtained across the whole film with a step size of 0.2 Å and a dwell time of 0.05 s/pixel. The EELS background was subtracted using a power-law function, and the multiple scattering effect was removed with a Fourier deconvolution method.

## III. RESULTS

### A. Atomic structure of $STO^5$-$SRO^n$-$STO^5$ (n = 1, 2) heterostructures

Bulk SRO crystallizes in an orthorhombic structure with space group *Pbnm* (no. 62) and a tilt pattern of $a^-a^-c^+$ at room temperature. The SRO grown on a $TiO_2$ terminated STO substrate undergoes a 0.5% compressive strain (bulk: $a_{STO}$ = 3.905 Å and $a_{pSRO}$ = 3.925 Å). The atomic structure of the $STO^5$-$SRO^n$-$STO^5$ (n = 1, 2) heterostructures on atomically flat region of the STO substrate was thoroughly investigated in our previous work [23]. Electric and magneto-transport measurements demonstrate that $STO^5$-$SRO^1$-$STO^5$ is insulating and non-FM, whereas $STO^5$-$SRO^2$-$STO^5$ is FM metallic with a Curie temperature of ∼128 K. The SRO-STO heterostructures are fully strained to the STO substrate; hence, the SRO layers exhibit slightly increased out-of-plane lattice parameters compared to the value for bulk SRO. Due to connectivity with $TiO_6$ octahedra in cubic STO, the ultrathin 2 u.c.- and monolayer SRO display tetragonal symmetry without $RuO_6$ tilt. Therefore, the $STO^5$-$SRO^n$-$STO^5$ heterostructures provide an ideal platform, without defects such as dislocations and out-of-phase domains, for studying substrate-surface step propagation and its effect on the microstructures of heterostructures.

**B. Two types of steps**

Throughout the entire film, the substrate step terraces are on a micrometer scale, similar to the AFM observations mentioned later. However, in order to microscopically reveal the step effects on SRO layers, we carefully searched for the region with bunched steps. Figure 1 displays the HAADF-STEM images from two representative step-propagating paths in the $STO^5$-$SRO^2$-$STO^5$-$SRO^1$ heterostructures. The intensity in the HAADF image is proportional to the atomic number ($Z^{2/1.7}$), and thus allows distinguishing the Ru (Z = 44), Sr (Z = 38), and Ti (Z = 22) columns in the images. The SRO-STO interfaces, marked by orange lines, highlight the step positions, allowing us to track step propagation within the heterostructure, which is not attainable in pure film. Terraces on the STO substrate surface, indicated by orange lines, are one unit-cell in height and range in width from 1 to 5 nm. Figure 1(a) presents steps propagating in the [011] direction from the higher terrace toward the lower terrace, referred to as type-I step, which is predominant in the film. As the film grows, the type-I step moves forward 1 u.c. in the upper layer with respect to the lower layer, leading to the propagation of the [011] direction. Some type-I steps proceed an extra 1 u.c. along the [010] direction, making the propagation slightly deviate from the [011] direction, as illustrated by arrows in Fig. 1(a). Figure 1(b) shows a type-II step that moves in the [0$\bar{1}$1] direction from the lower to upper terrace. In contrast to the type-I step, the type-II step moves backward 1 u.c. in the upper layer during the film growth process, resulting in propagation along the [0$\bar{1}$1] direction. Some type-II steps proceed one fewer unit cell in the [0$\bar{1}$0] direction, making the propagation slightly deviate from the [0$\bar{1}$1] direction.

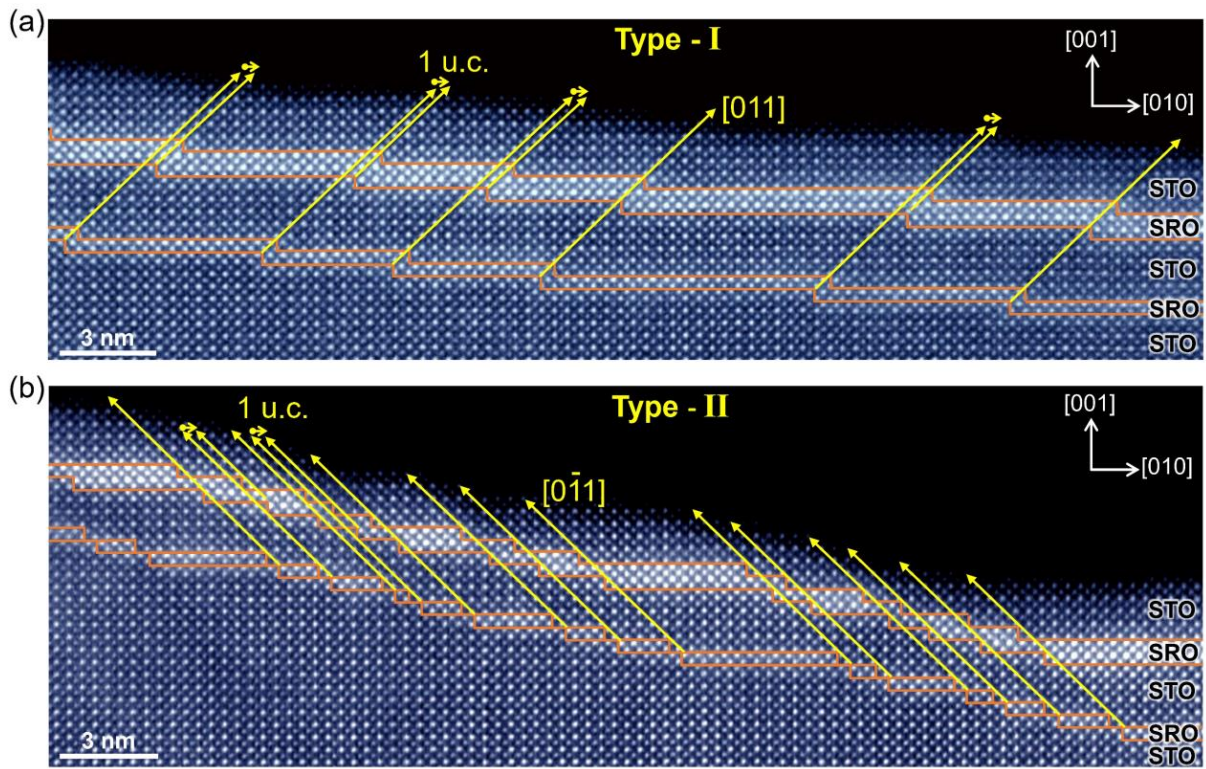

**FIG. 1.** Cross-sectional STEM image of a $STO^5$-$SRO^2$-$STO^5$-$SRO^1$ heterostructure on (001) STO substrate taken along the [100] direction, displaying two propagation modes of the 1 u.c.- height step within the heterostructure. (a) Type-I step propagating along the [011] direction from the upper toward lower terrace. (b) Type-II step spreading in the [0$\bar{1}$1] direction from the lower to upper terrace. The orange line marks $TiO_2$-SrO interfaces between substrate-heterostructure and SRO-STO layers in the heterostructure. Yellow arrows indicate the step propagation direction in the heterostructure. The 1 u.c. deviation in step propagation toward lower terrace is denoted by short yellow arrows.

We performed STEM-EELS mapping to study the elemental distribution in the heterostructure. Figure 2 displays Ti elemental map across the STO$^5$-SRO$^2$-STO$^5$-SRO$^1$ heterostructure, in which the SRO layers start with SrO layers and terminate with RuO$_6$ layers. The bottom SRO layers are 1 u.c. thick, and approximately 30% of Ru dopants into the TiO$_2$- terminated layer of STO substrate. As indicated by the arrows in Fig. 2b, the Ti columns show darker intensities in the elemental map and higher intensities in the Z-contrast HAADF image [see Fig. 2(a)], indicating the presence of Ru dopants in these Ti columns. In addition, the Ru columns at step edges exhibit lower intensities in the HAADF image of Fig. 2(a), suggesting Ti diffusion into the Ru columns. Such Ti-Ru intermixing primarily occurs within a unit cell located at the step edge. This observation indicates that the heterostructure prefers to grow from the step edge, even employing the layer-by-layer growth mode [23]. It provides direct evidence that the step edges are preferential nucleation sites because of more coordination [9]. The EELS elemental mapping is consistent with the HAADF observations, demonstrating that the intensity of the Z-contrast HAADF image allows us to locate the steps and Ru-Ti intermixing columns in examining the atomic structure of the steps described below.

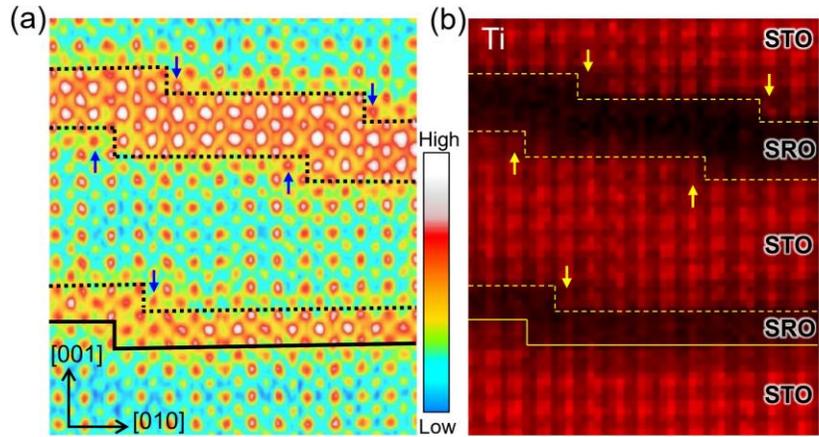

**FIG. 2.** Chemical intermixing at the step edges in STO$^5$-SRO$^2$-STO$^5$-SRO$^1$ observed by STEM-EELS. (a) False-color HAADF image and (b) corresponding EELS elemental map extracted from Ti-L edge. Solid orange lines denote the substrate-heterostructure interface, highlighting 1 u.c.- height step on substrate-surface. Dashed orange lines mark the TiO$_2$-SrO interfaces, highlighting step locations within the heterostructure. Arrows indicate Ti columns with Ru dopants at the step edges.

**C. Step effect on the sandwiched ultrathin SRO heterostructures**

Detailed analysis of atomic-resolved HAADF-STEM images (Fig. 3 and Fig. 4) reveals that these two types of steps induce lateral inhomogeneity within the heterostructure and affect the thickness of the SRO layers in opposite ways. The type-I [011] step moves forward in the subsequent layer, adding an extra unit cell to the SRO layers in the film growth direction between the step edges [see Fig. (3a)]. As indicated by the blue rectangle depicted in Fig. 3(b) and 3(c), the top 2 u.c.- SRO layer contains 3 u.c.- block while the bottom 1 u.c. SRO layer has 2 u.c.- block. The region in the 3 u.c.- SRO layer, located between the step edges, changed to 4 u.c. in the STO$^5$-SRO$^3$-STO$^5$-SRO$^1$ heterostructure (see Fig. A1 in Appendix), exhibiting the same tendency. On the other hand, the type-II [0$\bar{1}$1] step removes an SRO unit cell between the steps in the [001] direction as it moves backward in the subsequent layer. Figure 4 shows the top 2 u.c.- SRO block reduced to 1 u.c., and the bottom 1 u.c.- SRO block became discontinuous. Furthermore, Ti-Ru intermixing was observed at the step edges of STO-SRO interfaces.

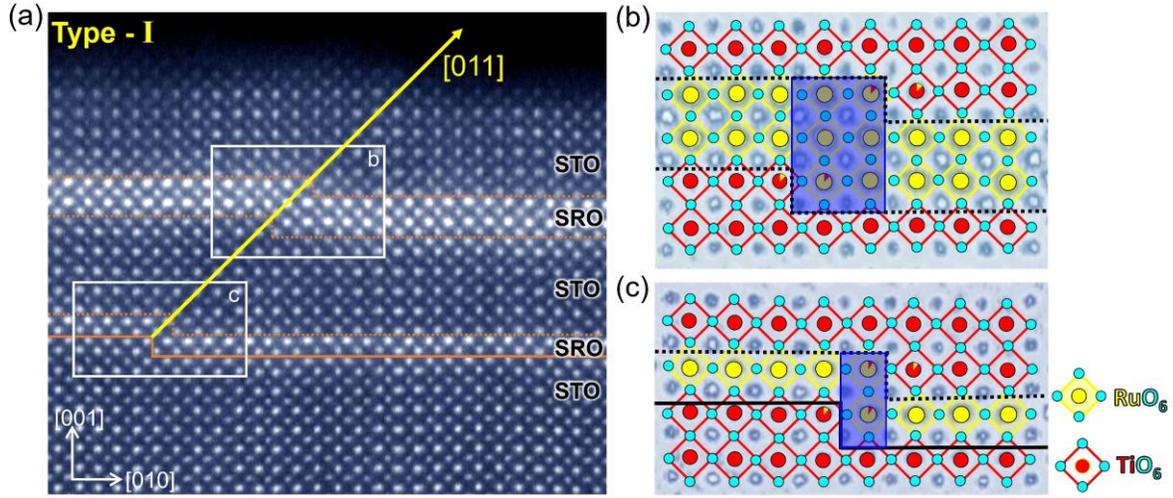

**FIG. 3.** Atomic arrangement of the Type-I step in $STO^5$-$SRO^2$-$STO^5$-$SRO^1$. (a) HAADF image of the Type-I step. Enlarged-view of steps in the (b) top 2 u.c.- and (c) bottom 1 u.c.- SRO layers sandwiched by STO. Crystal structure models are superimposed. The Ru-Ti intermixing marked by the yellow Ru-atoms with fractional occupation of Ti-atoms (red) is estimated from the intensity in the HAADF image. The STO substrate-heterostructure interface and $TiO_2$-SrO of STO and SRO layers are denoted by solid and dashed orange lines, respectively. Blue rectangles highlight the inhomogeneous regions in SRO layers between the steps.

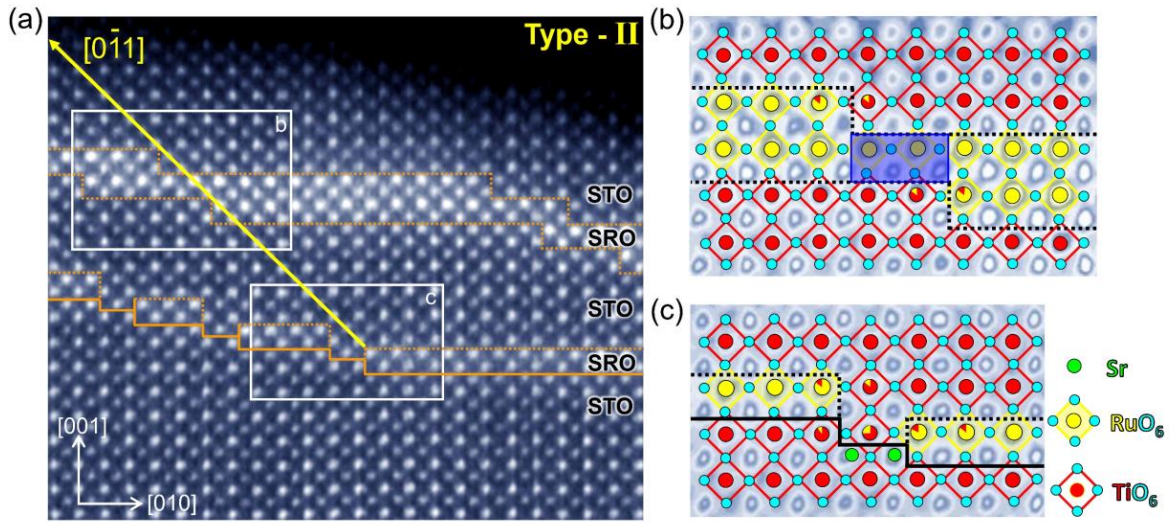

**FIG. 4.** Structure feature of the Type-II step in $STO^5$-$SRO^2$-$STO^5$-$SRO^1$. (a) HAADF image of the Type-II step. Enlarged image of steps in the (b) top 2 u.c.- and (c) bottom 1 u.c.- SRO layers from the rectangles in (a). Projected crystal structures are superimposed. Solid lines indicate the substrate-heterostructure interface, displaying SrO region along the substrate step edge. The dashed lines mark the $TiO_2$-SrO interfaces between the STO and SRO layers. Blue rectangles highlight the SRO blocks.

The schematic in Fig. 5 illustrates the structural features of these two types of steps in the STO-SRO heterostructure on the STO substrate with 1 u.c.-height steps. Within the region where the step passes through, the N u.c.- SRO layer changes in thickness. For the type-I step [see Fig. 5(a)], the layer thickness changed to N+1 u.c., forming an N × (N+1) block. Conversely, for the type-II step [see Fig. 5(b)], the SRO layer decreases from N to N-1 u.c., resulting in an N × (N-1) SRO block. For the type-II step, the in-plane continuity of the SRO layer will be destroyed when the SRO is reduced to 1 u.c. (N = 1) [see Fig. 5(d)], which would significantly affect the resistivity property of the monolayer SRO. Moreover, viewed perpendicularly to the interface, the $RuO_6$ rotation varies within the N × (N ± 1) block due to the one unit-cell difference. The in-plane inhomogeneity is a critical factor in understanding magnetic structures in ultrathin SRO films, as it relates to $RuO_6$ octahedron tilt and rotation.

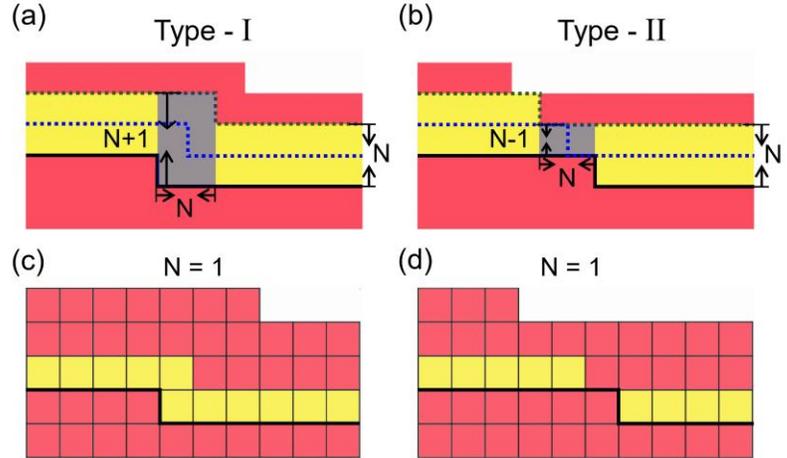

**FIG. 5.** Schematic illustration of formation process of the (a) type-I and (b) type-II steps on STO substrate with 1 u.c. height step. 'N' denotes the thickness of the SRO layer in unit cells. The monolayer SRO with N = 1 in (c) Type-I and (d) Type-II steps.

## IV. DISCUSSION

### A. Origin of the type-II $[0\bar{1}1]$ step

Since the type-II step causes discontinuities of monolayer SRO, its formation mechanism needs to be investigated. As illustrated in Fig. 5, the subsequent layer follows the step characteristics of the preceding layer in the film growth process, and hence, the step propagation mode is primarily determined by the first layer interfacing with the substrate. The 1 u.c. forward movement in the type-I step leads to a mimic of the substrate step, which is energetically favorable. Therefore, the type-I step was commonly observed. In contrast, the type-II step moves backward by 1 u.c. in the first layer, leaving a unit cell uncovered at the step edge. We will now delve into the driving force for the uncovered unit cell on the substrate. There are two primary hypotheses. The first is bunches of narrow steps on the STO substrate. The adatoms, which reach upper terrace edge, prefer to occupy the step corner of the narrow terrace, leaving the unit cell on the upper terrace edge uncovered. We observed a type-I step on a narrow terrace, and a type-II step was also observed on a wider terrace. Thus, our experimental results do not support this hypothesis. The other is that the uncovered unit cell is terminated with SrO, according to that SRO prefers to grow on the $TiO_2$-terminated layer rather than SrO on the termination mixed substrate [27]. Regrettably, in STEM images, it is not possible to discern the 1 u.c.- width SrO- terminated region on the substrate following the growth of the STO layer. However, a wider SrO- terminated region can be observed since it introduces vacancies in the STO layer.

HAADF images in Fig. 6 show a type-II step on the STO substrate with a 4 u.c.- width SrO- terminated region along the step edge. The SRO layers become discontinuous above the SrO- terminated region. Specifically, the top SRO layer [see Fig. 6(b)] exhibits partially discontinuous because most of Ru atoms are replaced by Ti in the 1u.c.- SRO block. The bottom single u.c. SRO layer is completely disconnected in the SrO- terminated region [see Fig. 6(c)], which is filled with STO. In addition, the darker intensity observed in the SrO- terminated region compared to the other parts of the film is caused by the presence of STO vacancies [see Fig. 5(a) and (c)]. The observation indicates that, during film growth process, the subsequent STO does not entirely fill the area above SrO-terminated region. The 1 u.c.- height step propagates in the $[0\bar{1}1]$ direction with respect to the lower terrace edge on the STO substrate. The region of thickness reduction extends to 3 u.c. in the top SRO layers, which is affected by both the thickness of SRO layers and the width of the SrO-terminated region on the substrate [see Fig. A2 in Appendix].

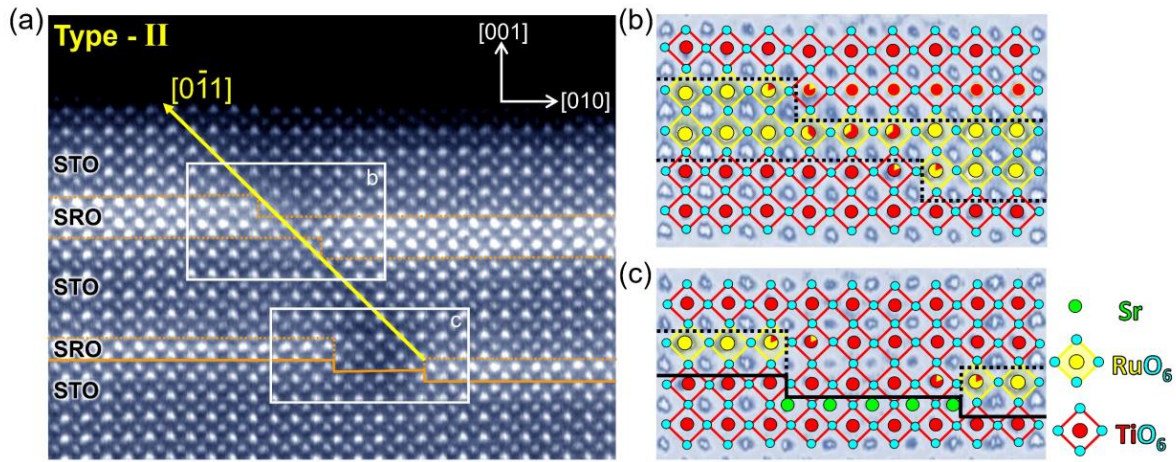

**FIG. 6.** Atomic structure of Type-II step in $STO^5$-$SRO^2$-$STO^5$-$SRO^1$ on STO substrate with 4 u.c. width SrO terminated region along step edge. (a) HAADF image showing the step propagating in the $[0\bar{1}1]$ direction. Enlarged image of steps in the (b) top 2 u.c.- SRO and (c) bottom 1 u.c.- SRO layers sandwiched by STO. The orange line marks the substrate-heterostructure interface, including 4 u.c.- width SrO terminated region. The dark intensity above non-SRO covered SrO terminated region is caused by STO vacancies in STO layer.

## B. Step effect on physical properties

Before discussing the effect of steps on physical properties, it is crucial to address the following question: Is the SrO terminated region decorated at step edges a common feature on the $TiO_2$-terminated STO substrate? BHF solution etching for 30 s in our case is proposed to be the optimal condition for removing the SrO- terminated region along the step, while a longer etching time would result in the formation of holes on the substrate [26]. Unfortunately, the optimal etching time varies depending on the quality of STO substrates [26]. The surface of STO presented in Fig. 7 is typically obtained by sonicating the substrate in deionized water for 30 minutes, then etching it with BHF for 36 s, and then annealing at 930 °C for half an hour in an oxidizing atmosphere. Figures 7(a) and (c) show the surfaces after these treatments. The surface of the STO is composed of steps and atomically flat large terraces. The height between adjacent large terraces is 3.9 Å, i.e., one STO unit cell high. All these large terraces consist of $TiO_2$ [5], while small terraces located at the edge of the steps are also observed. According to the line profile displayed in Fig. 7(b), these small terraces are 10-20 nm wide and ~2.0 Å high, which equals to the height of half STO unit cell. This suggests that these small terraces along the step edges are consisted of SrO layers.

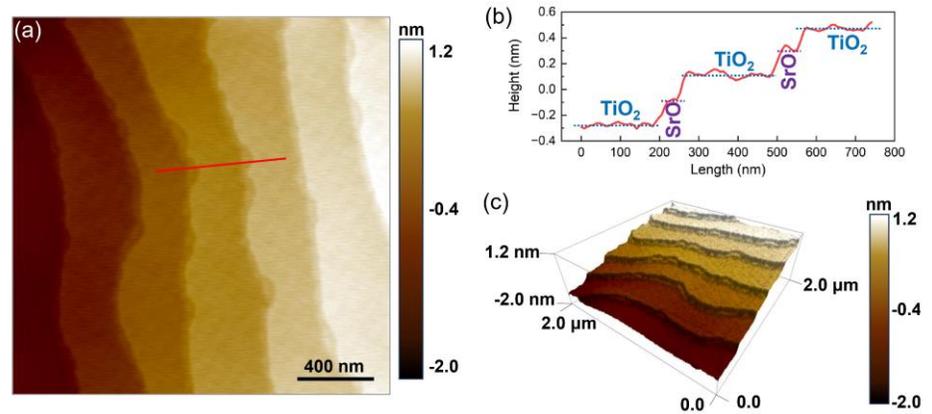

**FIG. 7.** (a) A typical AFM image of STO substrate with large $TiO_2$-terminated surface. (b) A line profile [marked as the red line in (a)]. (c) The corresponding 3D view of the STO surface displayed in (a).

To see the structural effect of the SrO layer decorated at the steps of $TiO_2$-terminated STO surface, we grew a 5 u.c. STO-capped SRO film on STO substrate and examined it with STEM. As shown in Fig. 8, the film is discontinuous above the wider SrO-terminated regions along the step edge of the $TiO_2$-terminated surface. The capped STO layers filled the bare SrO-terminated region, resulting in grooves on the film surface. The groove is wider than the SrO-terminated region on the substrate because the SRO layers grow in opposite directions.

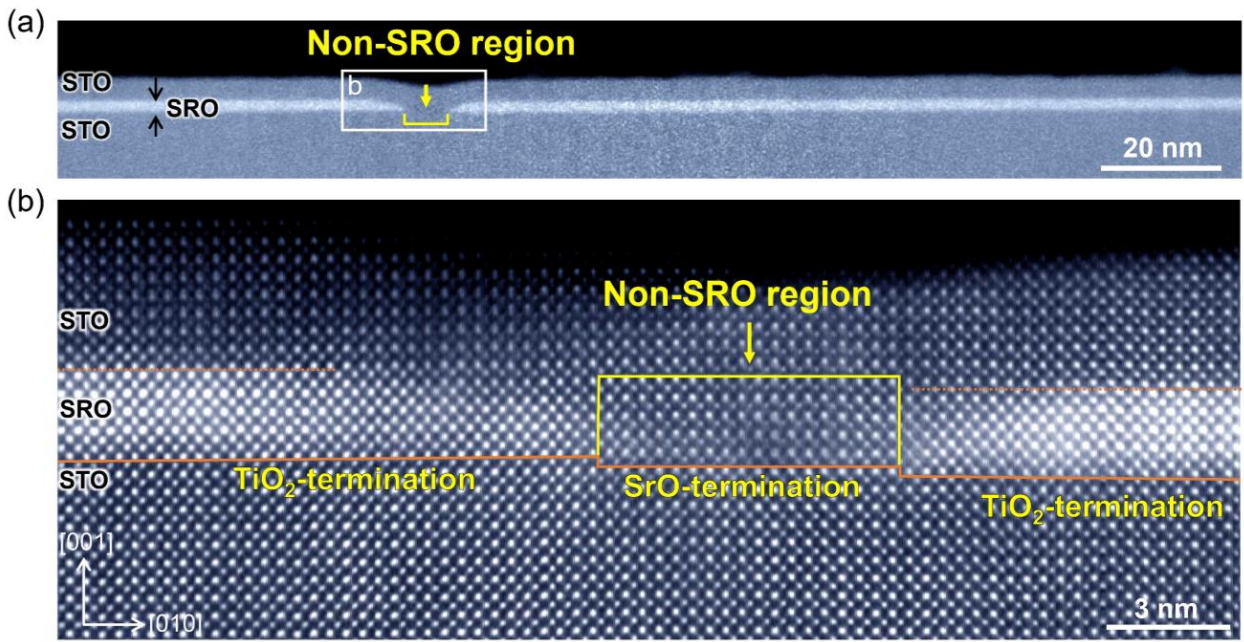

**FIG. 8.** Discontinuous SRO thin film induced by a SrO- terminated region along the 1 u.c.- height step on $TiO_2$-terminated STO substrate. (a) Low-magnification HAADF image of STO capped 5 u.c.- SRO film on STO(001) substrate taken along the [100] direction. Black arrows mark SRO film with bright intensity. The yellow arrow signifies a non-SRO region with dark intensity and groove in the capped STO. (b) Atomic-resolved HAADF image from the rectangle region in (a). Solid orange lines mark the substrate-heterostructure interface, showing SrO terminated region at the step edge. The non-SRO region was filled with the capped STO.

The STO step edges with SrO-decorated regions can also introduce lateral discontinuity (or inhomogeneities due to the replacement of Ru with Ti) in the SRO films and heterostructures. As demonstrated in Fig. 4, the steps with 1 u.c.- width SrO decoration can lead to discontinuity of monolayer SRO. A wider SrO-decorated region would destroy the continuity of a thicker SRO film. Given that conductivity is measured on the grown films, the atomic-level discontinuity, capable of abruptly interrupting the pathway in electric transport, should be taken into account in the understanding of nonmetallic "dead" layer issue in ultrathin SRO films and superlattices. It is worth noting that wider SrO terminated regions along the step edges on the STO surface can be identified from the low magnification AFM and STM images, but the unit-cell width SrO terminated region cannot be distinguished unambiguously.

Single $TiO_2$-terminated STO substrate at atomic level is a mandatory condition to achieve high-quality termination selective ultrathin films and to reveal their physical properties. The aforementioned SRO/STO heterostructure is termination selective in the growth process, while the well-known $LaAlO_3$(LAO)/STO film is indeed termination dependent in the interface properties too. In the LAO/STO(001) case, the LaO-$TiO_2$ interface exhibits two-dimensional electron gas(2DEG) while the $AlO_2$-SrO interface is insulating. Excluding the SrO- terminated region at the step edges is essential to understand the contrary effect of step edges on the 2DEG [28,29], electron phase separation at low temperatures [30], as well as the longstanding "dead" layer issue in ultrathin oxide films in general.

## V. CONCLUSIONS

In conclusion, the cross-sectional STEM study of the $STO^5$-$SRO^2$-$STO^5$-$SRO^1$ heterostructure, grown on a $TiO_2$ terminated vicinal STO(001) substrate, reveals two distinct step propagation paths. The type-I [011] propagation step moves from the higher to lower terrace, leading to an extra SRO unit cell between the steps in the film growth direction. In contrast, the type-II [0$\bar{1}$1] propagation step removes one SRO unit cell in the film growth direction, resulting in the discontinuity of one monolayer SRO. Furthermore, SrO- terminated region along the $TiO_2$- terminated step edge is found to be the origin of the type-II step. A precise depiction of lateral inhomogeneities at the atomic scale would help greatly enhance our grasp of the intrinsic physical properties of ultrathin films, particularly in the case of superlattices.

## Acknowledgements

This work was primarily supported by the US Department of Energy (DOE) under Grant No. DOE DE-SC0002136. The electron microscopic work done at Brookhaven National Laboratory (BNL) was sponsored by the US DOE-BES, Materials Sciences and Engineering Division, under Contract No. DE-SC0012704. Part of this work was supported by the National Key R&D Program of China (No. 2022YFA1403000), the National Natural Science Foundation of China (No. 12304035) and the Strategic Priority Research Program of Chinese Academy of Sciences (XDB33000000).

# Appendix A: Atomic structure of the Type-I step in $STO^5$-$SRO^3$-$STO^5$-$SRO^1$ heterostructure

Figure A1 displays a HAADF-STEM image of the $STO^5$-$SRO^3$-$STO^5$-$SRO^1$ heterostructure on the TiO$_2$ terminated STO substrate with 1 u.c.- height steps. The step advances in the [110] direction, with 1 u.c. or 2 u.c. deviation toward the lower terrace. In the [001] film growth direction, the top 3 u.c.- SRO layer changed to 4 u.c. between the step edges, as indicated by the blue rectangle in Fig. A1(b).

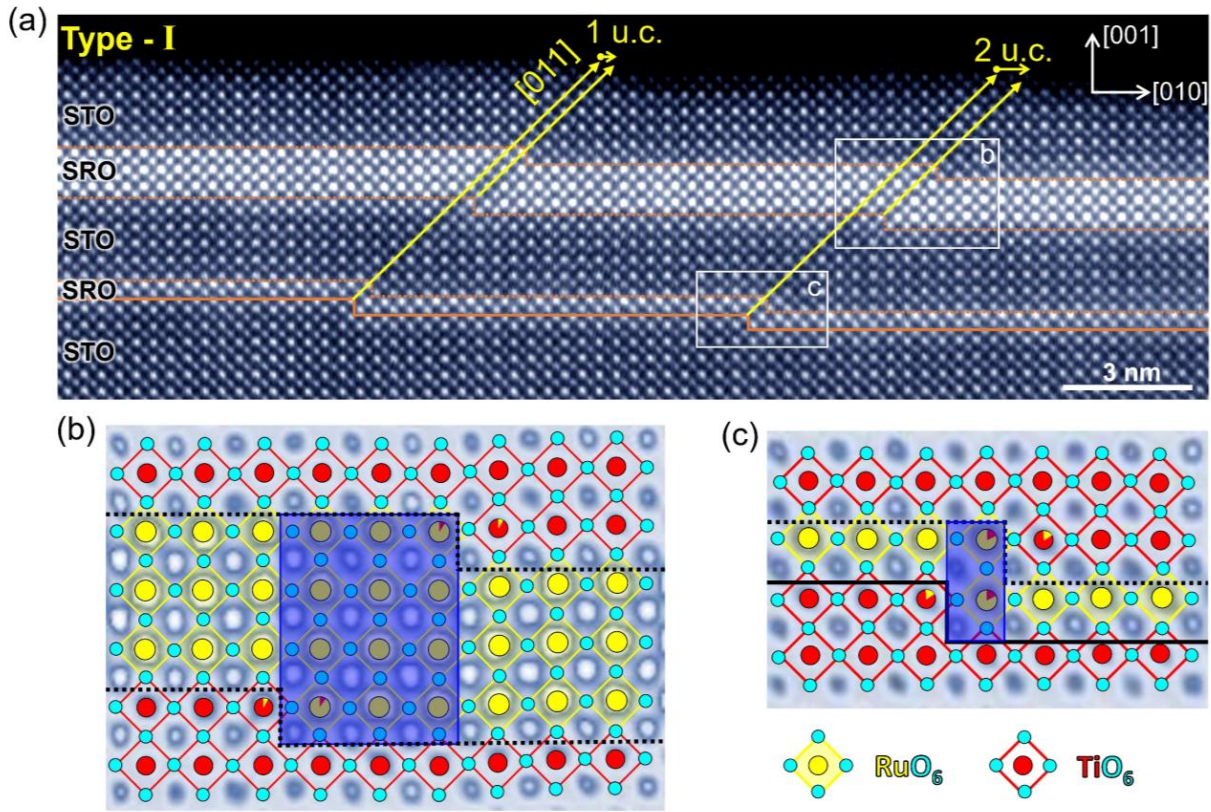

**FIG. A1.** Structure feature of the Type-I step in $STO^5$-$SRO^3$-$STO^5$-$SRO^1$ taken along the [100] direction. (a) HAADF image of the Type-I step. Enlarged images of steps within the (b) top 3 u.c.- and (c) bottom 1 u.c.- SRO layers from the rectangles in (a). Projected crystal structures are superimposed. Solid lines indicate the substrate-heterostructure interface. Dashed lines mark the TiO$_2$-SrO interfaces within STO-SRO heterostructure. Blue rectangles highlight the SRO blocks with changed thickness.

# Appendix B: Atomic structure of type-II step in $STO^5$-$SRO^3$-$STO^5$-$SRO^1$ heterostructure

Figure A2 displays the type-II step in the $STO^5$-$SRO^3$-$STO^5$-$SRO^1$ heterostructure. The orange lines highlight a 2 u.c.- height step on the $TiO_2$ terminated STO substrate, featuring a 5 u.c.- width SrO- terminated region at the step edge. The 2 u.c.- height step grows in the [1-10] direction with respect to the lower terrace, as denoted by yellow arrows in Fig. A2(a). The bottom 1 u.c.- SRO layer is discontinuous due to the absence of SRO coverage on the SrO- terminated region. The dark intensity above the SrO- terminated region indicates that subsequent growth STO does not completely fill this region and introduce STO vacancies. The thickness of the top 3 u.c.- SRO layer changes to 2 u.c. and 1 u.c. between the steps, as illustrated by the crystal structure models in Fig. A2(b).

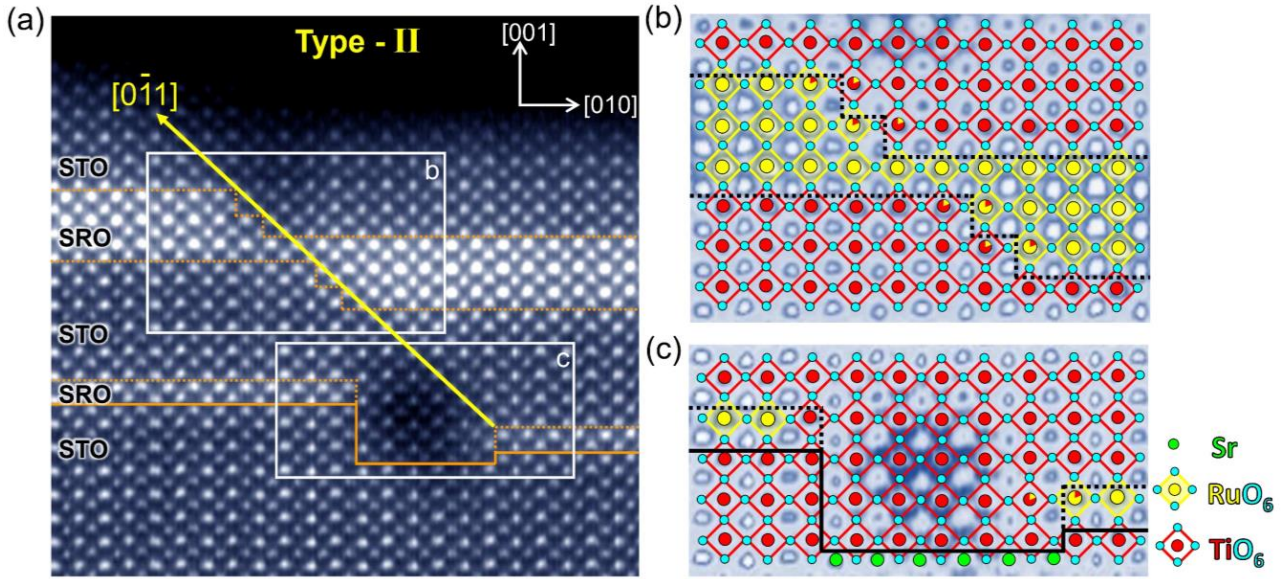

**FIG. A2.** Type-II step in $STO^5$-$SRO^3$-$STO^5$-$SRO^1$ heterostructure on the $TiO_2$ terminated STO substrate with SrO-terminated region along a 2 u.c.- height step edge. (a) HAADF image showing the step propagation in the $[0\bar{1}1]$ direction. Enlarged view of the areas indicated by rectangles in (a) showing steps in the (b) top 3 u.c.- and (c) bottom 1 u.c.- SRO layers. In (c), solid black line marks the STO substrate surface with 5 u.c.- width SrO-terminated region at the step. The dark intensity above the SrO terminated region is induced by STO vacancies.